\documentclass[letterpaper, 11pt]{article}
\usepackage{appendix}
\usepackage{multirow}
\usepackage{algpseudocode}
\usepackage{amssymb, amsmath, amsthm, amsfonts}
\usepackage{algorithm}
 
\newcommand{\commentOut}[1]{}
 
\usepackage{graphicx}

\usepackage{subfigure}
\usepackage{amsmath}
\usepackage{amssymb}
\usepackage{mdwlist}

\usepackage{hyperref}

\usepackage{xspace}
\usepackage{setspace}

\def\noflash#1{\setbox0=\hbox{#1}\hbox to 1\wd0{\hfill}}

\newcommand{\scripte}{\mathcal{E}}
\newcommand{\scriptv}{\mathcal{V}}

\begin{document}

\title{Recent Results on Fault-Tolerant Consensus \\in Message-Passing Networks\thanks{A shorter version of the survey is published in SIROCCO 2016.}}


%
%
\author{Lewis Tseng~\\~\\ltseng3@illinois.edu}

\date{May 2016\footnote{Revised on August 2016 to improve the presentation.}}
\maketitle

\begin{abstract}
	Fault-tolerant consensus has been studied extensively in the literature, because it is one of the most important distributed primitives and has wide applications in practice.
	This paper surveys important results on fault-tolerant consensus in message-passing networks, and the focus is on results from the past decade. 
	Particularly, we categorize the results into two groups: new problem formulations and practical applications. 
	In the first part, we discuss new ways to define the consensus problem, which includes larger input domains, link fault models, different network models $\dots $etc, and briefly discuss the important techniques.
	In the second part, we focus on Crash Fault-Tolerant (CFT) systems that use Paxos or Raft, and Byzantine Fault-Tolerant (BFT) systems. 
	We also discuss Bitcoin, which can be related to solving Byzantine consensus in anonymous systems, and compare Bitcoin with BFT systems and Byzantine consensus.
	
\end{abstract}

\section{Introduction}

Fault-tolerant {\em consensus} has received significant attentions over the past three decades \cite{welch_book,AA_nancy} since the seminal work by Lamport, Shostak, and Pease \cite{lamport_agreement,lamport_agreement2} --
some important results include solving consensus efficiently and identifying time and communication complexity under different models -- please refer to \cite{welch_book,AA_nancy,Raynal_book,Raynal_book} for these fundamental results.
In this paper, we survey recent efforts on fault-tolerant consensus in message-passing networks, with the focus on results from the past decade. 
References \cite{Byz_survey,Raynal_guide,Cachin_SMR} presented early surveys on the topic.
To complement theses prior works, we present this survey from two new angles:

\begin{itemize}
	\item {\em Exploration of New Problem Formulations}: Lots of different consensus problems have been introduced in the past ten years for achieving more complicated tasks and accommodating different system and network requirements. New problem formulations include enriched correctness properties, different fault models, different communication networks, and different input/output domains. For this part, we focus on the comparison of recently proposed problem formulations and relevant techniques.
	
	\item {\em Exploration of Practical Applications}: 
	Consensus has been applied in many practical systems. Here, we focus on three types of applications: (i) crash-tolerant consensus algorithms (mainly Paxos \cite{Paxos_1988} and Raft \cite{Raft}) and their applications in real-world systems, (ii) PBFT (Practical Byzantine Fault-Tolerance) \cite{PBFT_1999} and subsequent works on improving PBFT, and (iii) Bitcoin \cite{bitcoin_web} and its relationships with Byzantine consensus and BFT (Byzantine Fault-Tolerance) systems.
\end{itemize}

\paragraph{Classic Definitions of Fault-tolerant Consensus}
\label{s:classic}

~

We consider the consensus problem in a point-to-point message-passing network, which is modeled as an undirected graph.
Without specifically mentioning, the communication network is assumed to be \textit{complete} in this survey, i.e., each pair of nodes can communicate with each other directly.
In the fault-tolerant consensus problem \cite{welch_book,AA_nancy}, each node is given an {\em input}, and after a finite amount of time, each fault-free node should produce an {\em output} -- consensus algorithms should satisfy the {\em termination} property.
Additionally, the algorithms should also satisfy appropriate \textit{validity} and {\em agreement} conditions. 
There are three main categories of consensus problems regarding different agreement properties:

\begin{itemize}
	\item {\em Exact } \cite{lamport_agreement,Paxos_1988}: fault-free nodes have to agree on exactly the \textit{same} output.
	\item {\em Approximate }  \cite{AA_Dolev_1986,AA_Fekete_aoptimal}: fault-free nodes have to agree on ``roughly'' the \textit{same} output -- the difference between outputs at any pair of fault-free nodes is bounded by a given constant $\epsilon~~(\epsilon > 0)$ of each other.
	\item {\em $k$-set }  \cite{k-set_oldest,k-set_async}: the number of distinct outputs at fault-free nodes is $\leq k$.
\end{itemize}

Validity property is also required for consensus algorithms to produce meaningful outputs, since the property defines the acceptable relationship between inputs and output(s). Typical validity includes: (i) \textit{strong validity}: output must be an input at some fault-free node, (ii) \textit{weak validity}: if all fault-free nodes have the same input $v$, then $v$ is the output, and (iii) \textit{validity} (for approximate consensus): output must be bounded by the inputs at fault-free nodes.
A consensus algorithm is said to be correct if it satisfies termination, agreement and validity properties given that enough number of nodes are fault-free throughout the execution of the algorithm. In this paper, we focus on three types of node failures -- Byzantine, crash, and omission faults. The only exception is Section \ref{s:link} where we discuss results on link faults.

The other key component of the consensus problem definition is {\em system synchrony}, i.e., a model specifying the relative speed of nodes and the network delay.
There are also three main categories \cite{welch_book,AA_nancy,partial_synchrony,minimal_synchrony}:

\begin{itemize}
	\item \textit{Synchronous}: each node proceeds in a lock-step fashion, and there is a known upper bound on the message delay.
	\item \textit{Partially synchronous}: there exists a {\em partially synchronous} period from time to time. In such a period, fault-free nodes and the network stabilize and behave (more) synchronously.\footnote{Note that there are also other definitions of partial synchrony. We choose this particular definition, since many BFT systems only satisfy liveness under this particular definition. Please refer to \cite{partial_synchrony,partial_synchrony_aguilera} for more models on partial synchrony.}
	\item \textit{Asynchronous}: no known bound exists on nodes' processing speed or the message delay. 
\end{itemize}

\paragraph{Outline}

~

Section \ref{s:new} discusses works that defined new consensus problems which either assumed variants of aforementioned properties or introduced enriched correctness properties. The main purpose is to give a big picture on the problem space that have been explored in the literature.
In Section \ref{s:bft}, we discuss recent efforts that bring consensus algorithms to practical systems. Consensus is an important primitive that has wide applications such as state-machine replication (SMR) \cite{RSM}, and distributed storage. We address three main applications: (i) crash fault-tolerant systems that used variants of Paxos \cite{Paxos_1988,Paxos_simple} and Raft \cite{Raft}, (ii) Byzantine fault-tolerant (BFT) systems, and (iii) Bitcoin \cite{bitcoin_Satoshi,bitcoin_princeton}, a popular cryptocurrency. For part (ii), we discuss several techniques on improving the performance, including speculative execution, execution/agreement separation, and hardware-based solution $\dots$ etc. For part (iii), we focus on the comparison of Bitcoin and Byzantine consensus and BFT systems.
In Section \ref{s:future}, we conclude the survey, and present some interesting future research directions.

\section{Exploration of New Problem Formulations}
\label{s:new}

Researchers have generalized the consensus problem from the classic definitions presented in Section \ref{s:classic}. We categorize these efforts into four groups: (i) input/output domain, (ii) communication network and synchrony assumptions, (iii) link fault models, and (iv) enriched correctness properties, such as early-stopping and one-step properties. In this section (with the exception in Section \ref{s:link} when we discuss link fault models), we assume that there are $n$ nodes in the system, and up to $f$ of them may become Byzantine faulty or crash. Byzantine faulty nodes may have an arbitrary behavior.

\subsection{Input/Output Domain}

\paragraph{Multi-Valued Consensus}


In the original \textit{exact} Byzantine consensus problem \cite{lamport_agreement,lamport_agreement2}, both input and output are binary values. 
Later, references \cite{Multi-valued_lynch,Multivalue_Coan} proposed the multi-valued version in which input may take more than two \textit{real} values.
Recently, multi-valued consensus received renewed attentions and researchers proposed algorithms that achieve asymptotically optimal communication complexity (number of bits transmitted) in both synchronous and asynchronous systems. Surprisingly, for an $L$-bit input, these algorithms achieve asymptotic communication complexity of $O(nL)$ bits when $L$ is large enough.

In synchronous systems,
Fitzi and Hirt proposed a Byzantine multi-valued algorithm with small error probability \cite{Multi-valued_Hirt}. Their algorithm is based on the reduction technique and has the following steps: (i) hash the inputs to much smaller values using universal hash function, (ii) apply (classic) Byzantine consensus algorithm using these hash values as inputs, and (iii) achieve consensus by obtaining the input value from nodes that have the same hash values (if there is enough number of such nodes) \cite{Multi-valued_Hirt}. Later, Liang and Vaidya combined a different reduction technique (that divides an input into a large number of small values)
with novel coding technique to construct an error-free algorithm in synchronous systems \cite{Multi-valued_liang}. One key contribution is to introduce a lightweight fault detection (or fault diagnosis) mechanism using coding \cite{Multi-valued_liang}. Their coding-based fault diagnosis is efficient because the inputs are divided into batches of small values. In each batch, either consensus (on the small value of this batch) can be achieved with small communication complexity or some faulty nodes will be identified. Once all faulty nodes are identified, then consensus on the remaining batches becomes trivial. Since number of faulty node is bounded, consensus on most batches can be achieved with small communication complexity \cite{Multi-valued_liang}.

Subsequently, variants of reduction technique were applied to solve consensus problems with large inputs in asynchronous systems. References \cite{Multi-valued_11,Multi-valued_10} provided multi-valued algorithms with small error probability. 
Afterwards, Patra improved the results and proposed an error-free algorithm \cite{Multi-valued_async}.
These algorithms terminate with overwhelming probability; however, the expected time complexity is large because these algorithms first divide inputs to small batches and achieve consensus on each batch using variants of fault diagnosis mechanisms. 

Typically, to achieve optimal communication complexity, the number of batches is in the same order of $L$.
Consequently, the number of messages is large, since by assumption, $L$ is a large value.
Instead of achieving optimal bits, Most\'{e}faoui and Raynal focused on a different goal -- minimizing number of messages in asynchronous systems \cite{Raynal_SIROCCO15,Raynal_Acta}. 
Their algorithm relies on two new all-to-all communication abstractions, which have an $O(n^2)$ message complexity (i.e., $O(n^2L)$ bits) and a constant time complexity.  The first one allows the fault-free to reduce the number of input values to a small constant $c$, which ranges from $3$ to $6$ depends on the bound on the number of faulty nodes.  The second abstraction allows each fault-free nodes to obtain a set of inputs such that, if the set at a fault-free node contains a single value, then this value belongs to the set of any fault-free process. The algorithm in \cite{Raynal_SIROCCO15,Raynal_Acta} consists of four phases such that (i) nodes exchange input values in the first three phases with the first phase based on the first communication abstraction, and the two subsequent phases based on the second, and (ii) nodes use binary consensus in the final phase to determine whether it is safe to agree on the value learned from phase 3.
Recently, Most\'{e}faoui and Raynal proposed a new all-to-all communication abstraction, validated broadcast, and showed how to used validated broadcast to solve the Byzantine consensus problem in asynchronous systems \cite{Raynal_ITBC}. The resulting algorithm has the intrusion-tolerant property: if all the faulty nodes propose the same value $v$, while no fault-free nodes proposes it, then $v$ cannot be the output.

Multi-valued consensus has also been studied under the crash fault model in which nodes may suffer fail-stop failures; otherwise, they follow the algorithm specification.
Most\'{e}faoui et al. proposed multi-valued consensus algorithms in both synchronous and asynchronous systems \cite{Multi-valued_crash}.
Later, Zhang and Chen proposed a more efficient multi-valued consensus algorithm in asynchronous systems \cite{Multi-valued_cost}.

\paragraph{High-Dimensional Input/Output}

In the Byzantine vector consensus (or multi-dimensional consensus) \cite{herlihy_multi-dimension_AA,Vaidya_BVC}, each node is given a $d$-dimensional vector of reals as its input ($d \geq 1$), and the output is also a $d$-dimensional vector.
In complete networks, the recent papers by Mendes and Herlihy \cite{herlihy_multi-dimension_AA}
and Vaidya and Garg \cite{Vaidya_BVC} addressed approximate vector consensus in the presence of Byzantine faults.
These papers
yielded lower bounds on the number of nodes, and algorithms with optimal resilience in asynchronous \cite{herlihy_multi-dimension_AA,Vaidya_BVC} as well as synchronous systems \cite{Vaidya_BVC}.
The algorithms in \cite{herlihy_multi-dimension_AA,Vaidya_BVC} are generalizations of the optimal iterative approximate Byzantine consensus for scalar inputs in asynchronous systems \cite{abraham_04_3t+1_async}. The algorithms in \cite{herlihy_multi-dimension_AA,Vaidya_BVC} require sub-routines for geometric computation in the $d$-dimensional space to obtain the local state in each iteration; whereas, a simple average operation suffices when $d=1$ \cite{abraham_04_3t+1_async}.
These two papers \cite{herlihy_multi-dimension_AA} and \cite{Vaidya_BVC} independently addressed the same problem, and developed different algorithms -- mainly on different geometric computation techniques, which also result in different proofs.

Subsequent work by Vaidya \cite{Vaidya_incomplete} explored the approximate vector consensus problem in incomplete \textit{directed} graphs. Later, Tseng and Vaidya \cite{Tseng_podc14} proposed the convex hull consensus problem, in which fault-free nodes have to agree on ``largest possible'' polytope in the $d$-dimensional space that may not necessarily equal to a $d$-dimensional vector (a single point). The asynchronous algorithm in \cite{Tseng_podc14} bears some similarity to the ones in \cite{herlihy_multi-dimension_AA,Vaidya_BVC,abraham_04_3t+1_async}; however, Tseng and Vaidya used a different communication abstraction to achieve the ``largest possible'' polytope. Moreover, Tseng and Vaidya introduced a new proof technique to show the correctness of iterative consensus algorithms when the output is a polytope \cite{Tseng_podc14}.


\subsection{Communication Network and Synchrony}


The fault-tolerant consensus problem has been studied extensively in complete networks (e.g., \cite{lamport_agreement,Paxos_1988,welch_book,AA_nancy,AA_Dolev_1986}) and in undirected networks (e.g., \cite{impossible_proof_lynch,dolev_82_BG}).
In these works, any pair of nodes can communicate with each other reliably either directly or via at least $2f+1$ node-disjoint paths (for Byzantine faults) or $f+1$ node-disjoint paths (for crash faults). Recently, researchers revisited assumptions on the communication network and enriched the problem space in three main directions: directed graphs, dynamic graphs, and partial synchrony.



\paragraph{Directed Graphs}

Researchers started to explore various consensus problems in arbitrary directed graphs, i.e., two pairs of nodes may not share a bi-directional communication channel, and not every pair of nodes may be able to communicate with each other directly or indirectly. 
Significant efforts have also been devoted on \textit{iterative} algorithms in incomplete graphs. In iterative algorithms, (i) nodes proceed in iterations; (ii) the computation of new state at each node is based only on local information, i.e., node’s own state and states from neighboring nodes; and (iii) after each iteration of the algorithm, the state of each fault-free node must remain in the convex hull of the states of the fault-free nodes at the end of the previous iteration. Vaidya et al. \cite{vaidya_PODC12} proved {\em tight} conditions for achieving approximate Byzantine consensus in synchronous and asynchronous systems using \textit{iterative} algorithms. The tight condition for achieving approximate crash-tolerant consensus using iterative algorithms in asynchronous systems was also proved in \cite{Tseng_thesis}.

A more restricted fault model -- called ``malicious'' fault model -- in which the faulty nodes are
restricted
to sending identical messages to their neighbors has also been explored extensively, e.g., \cite{Leblanc_HSCC_1,Leblanc_HSCC_2,Sundaram_ACC,Sundaram_journal}. 
LeBlanc and Koutsoukos \cite{Leblanc_HSCC_1} addressed a continuous time version of the consensus problem with malicious faults in complete graphs. 
LeBlanc et al. \cite{Sundaram_journal} have obtained {\em tight} necessary
and sufficient conditions for tolerating up to $f$ faults in the network.

The aforementioned approximate algorithms (e.g., \cite{vaidya_PODC12,lili_multihop,Sundaram_journal}) are generalizations of the iterative approximate consensus algorithm in complete network \cite{AA_Dolev_1986,AA_Fekete_aoptimal}. However, to accommodate directed links, the proofs are more involved. Particularly, for the sufficiency part, one has to prove that all fault-free nodes must be able to receive a non-trivial amount of a state at some fault-free node in finite number of iterations. 
The necessity proofs in the work on directed graphs (e.g., \cite{vaidya_PODC12,Sundaram_journal}) are generalizations of the indistinguishability proof \cite{impossibility_book,impossible_proof_lynch}. The main contributions were to identify how faulty nodes can block the information flow so that (i) fault-free nodes can be divided into several groups, and (ii) there exists a certain faulty behavior such that different groups of fault-free nodes have to agree on different outputs.

There were also works on using general algorithms to achieve consensus -- An algorithm is \textit{general} if nodes are allowed to have topology knowledge and the ability to route messages (send and receive messages using multiple node-disjoint paths). Furthermore, unlike iterative algorithms (e.g., \cite{AA_Dolev_1986,abraham_04_3t+1_async}), the state maintained at each node in general algorithms is not constrained to a single value.
Tseng and Vaidya \cite{Tseng_podc2015} proved \textit{tight} necessary and sufficient conditions on the underlying communication graphs for achieving (i) exact crash-tolerant consensus in synchronous systems, (ii) approximate crash-tolerant consensus in asynchronous systems, and (iii) exact Byzantine consensus in synchronous systems using \textit{general} algorithms.
Lili and Vaidya \cite{lili_multihop} proved tight conditions for achieving approximate Byzantine consensus using general algorithms. The exact consensus algorithms in \cite{Tseng_podc2015} also require that some information has to be propagated to all fault-free nodes even if some nodes may fail. Generally, the algorithms in \cite{Tseng_podc2015} proceed in phases such that in each phase, a group of nodes try to send information to the remaining nodes. The algorithms are designed to maintain validity at all time. Additionally, if no failure occurs in a phase, then agreement can be achieved. The algorithm in \cite{lili_multihop} can be viewed as an extension of the iterative algorithm in \cite{vaidya_PODC12}, which utilized the routing information to tolerate more failures than the algorithm in \cite{vaidya_PODC12} does.

\paragraph{Dynamic Graphs}

Researchers have also explored the consensus problem in directed dynamic networks \cite{k-set_dynamic,agreement_dynamic,approximate_consensus_dynamic,Charron-Bost,k-set_direct}, where communication network changes over time. For synchronous systems,
Charron-Bost et al. \cite{approximate_consensus_dynamic,Charron-Bost} solved {\em approximate} crash-tolerant consensus in directed dynamic networks
using iterative algorithms.
In the asynchronous setting, Charron-Bost et al. \cite{approximate_consensus_dynamic,Charron-Bost} addressed approximate consensus
with crash faults in {\em complete} graphs. Roughly speaking, in the crash fault model, references \cite{approximate_consensus_dynamic,Charron-Bost} and references \cite{vaidya_PODC12,Tseng_podc2015} independently found out that it is necessary and sufficient to have a fault-free node that can reach every other fault-free node in the dynamic graphs and directed graphs, respectively.

References \cite{k-set_dynamic,k-set_direct,agreement_dynamic} considered the {\em message adversary}, which controls the communication pattern, i.e., the adversary has the power to specify the sets of communication graphs. Biely et al. studied the exact consensus problem \cite{agreement_dynamic} and $k$-set consensus problem \cite{k-set_dynamic,k-set_direct} in dynamic networks under the message adversary. All the nodes are assumed to be fault-free in \cite{k-set_dynamic,k-set_direct,agreement_dynamic}, and no message is tampered in message adversary model.

The algorithms in aforementioned papers share some similarity with their counter parts in complete graphs, e.g., \cite{AA_nancy,welch_book}. The main contributions of these papers are to identify concise definitions of dynamic graphs so that useful information can be propagated to enough number of nodes in the presence of faults or dynamic links. 

\paragraph{Unknown and Anonymous Networks} 

A network is \textit{unknown} if the set and number of participating nodes are previously unknown. Inspired by the observation that many self-organizing systems initially behave as unknown networks, researchers studied fault-tolerant consensus in unknown networks. The problem is named FT-CUP (Fault-Tolerant Consensus with Unknown Participants).
For crash faults,
Cavin et al. \cite{FT_CUP} proposes the FT-CUP problem, and identified necessary and sufficient conditions on the system composition and synchrony conditions in order to solve FT-CUP. The proposed algorithm is based on the usage of failure detectors.
Subsequently, Greve and Tixeuil \cite{FT_CUP_tradeoff} studied the tradeoff between \textit{knowledge connectivity} and synchrony condition. Knowledge connectivity represents each node's ``knowledge'' (or view) of the network, i.e., those nodes that are reachable. Their algorithm consists of two phases: (i) identify a set of nodes that share the same view of the network, and (ii) execute traditional consensus algorithm to reach consensus in the unknown networks. 
Alchieri et al. \cite{BFT-CUP_OPODIS} extended the problem to the Byzantine version, where some nodes may become Byzantine faulty. Their algorithm also consists of two phases: identifying enough nodes and then using traditional consensus algorithm among these nodes to solve the problem. To accommodate Byzantine behaviors, the algorithm is more complicated than the one in \cite{FT_CUP_tradeoff}, and the required knowledge connectivity is different from the ones in \cite{FT_CUP,FT_CUP_tradeoff}.
Note that the notion of knowledge connectivity is different from the connectivity in communication network (e.g., directed or dynamic networks discussed above). In \cite{FT_CUP,FT_CUP_tradeoff,BFT-CUP_OPODIS}, the underlying communication graph is assumed to be \textit{fully-connected} (i.e., complete network), and their algorithm does not work if the network is directed or dynamic.

There were also works on anonymous networks, in which the nodes do not have unique identity. Delporte-Gallet et al. \cite{Unknownanonymous_consensus} considered the problem of reaching consensus when nodes may crash in anonymous networks. Their algorithms emulate shared memory and solve consensus under different synchrony conditions. One main novelty is their ``leader election'' protocol that works in anonymous networks. Their leader election protocol is different from the traditional ones, because there may be multiple leaders; however, by carefully integrating the leader election protocol, the algorithm ensures that as long as multiple leaders behave identically, the consensus can be achieved \cite{Unknownanonymous_consensus}. For Byzantine consensus in synchronous anonymous networks, Okun and Barak \cite{anonymous_Okun} considered the case when nodes are able to distinguish messages received from different links (or ports).
In these two papers, the system is \textit{partially anonymous} in the sense that for a given message, a node is still able to learn the source of the message \cite{bitcoin_miller_anonymous}, even though the source may not have a distinct identity. In a \textit{completely anonymous} network, a node is not able to learn the source for any message, and there is no notion of links or ports. As a result, for any two received messages, a node cannot know whether these two messages are from the same source or from two different sources. Below, we discuss two works \cite{bitcoin_miller_anonymous,bitcoin_garay} that solved Byzantine consensus in completely anonymous network.
Inspired by Bitcoin \cite{bitcoin_Satoshi}, Miller and LaViola proposed a Byzantine consensus algorithm for asynchronous anonymous networks that uses moderately-hard puzzles \cite{bitcoin_miller_anonymous}. Such computational puzzles are useful in completely anonymous networks, because they do not require the knowledge of identities and prevent the (computationally bounded) adversary from gaining too much influence. The algorithm in \cite{bitcoin_miller_anonymous} is Monte Carlo algorithm in the sense that it violates the validity condition with small probability. Recently, Garay et al. \cite{bitcoin_garay} proposed an algorithm that satisfies validity with overwhelming probability. The algorithm was also based on Bitcoin \cite{bitcoin_Satoshi}.

\paragraph{Partial Synchrony}

Alistarh et al. \cite{set_partial_synchrony} considered $k$-set consensus in partially synchronous systems, and presented the asymptotically tight bound on the complexity of set agreement in such systems. 
Milosevic et al. \cite{transmission_fault} considered permanent and transient transmission faults in a variation of partially synchronous systems, and proved necessary and sufficient conditions on the number of nodes $n$ to tolerate permanent and transient transmission faults. Hamouma et al. \cite{Byzantine_consensus_synchronous_links} studied the consensus problem when only a few links may be synchronous throughout the execution of the algorithm.

Alistarh et al. \cite{consensus_smallest_window} addressed a fundamental question of partially synchronous systems: ``For how long does the system need to be synchronous to solve crash-tolerant consensus?'' The core idea of the algorithm in \cite{consensus_smallest_window} relies on two mechanisms (i) detect asynchrony, and (ii) determine when to update value safely (without violating the validity) based on asynchrony detection. 
Bouzid et al. \cite{minimal_synchrony} studied the problem from a different aspect -- how many eventually synchronous links are necessary for achieving consensus? They introduced a notion of eventual $\langle t+1 \rangle$bisource which characterizes the necessary and sufficient timing condition to solve consensus. This condition requires an existence of fault-free nodes such that it has an eventually synchronous incoming links from $f$ other fault-free nodes, and eventually synchronous outgoing links to $f$ other fault-free nodes. The proposed algorithm in \cite{minimal_synchrony} uses two novel components: a new all-to-all communication abstraction for fault-free nodes to eventually agree on a set of values, and an object to ensure that fault-free nodes eventually converge to a single value.

\subsection{Link Fault Model}
\label{s:link}


In addition to node failures,
significant efforts have also been devoted to the problem of achieving consensus in the presence of link failures \cite{HeardOf,Biely_hybrid,Santoro_link,Santoro_link2,impossible_link}.  Santoro and Widmayer proposed the {\em transient} Byzantine link failure model: a different set of links can be faulty at different time \cite{Santoro_link,Santoro_link2}. The nodes are assumed to be fault-free in the model. They characterized a necessary condition and a sufficient condition for undirected networks to achieve consensus in the transient link failure model; however, the necessary and sufficient conditions do not match: the necessary and sufficient conditions are specified in terms of node degree and edge-connectivity,\footnote{A graph $G=(\scriptv, \scripte)$ is said to be $k$-edge connected, if $G'=(\scriptv, \scripte - X)$ is connected for all $X \subseteq \scripte$ such that $|X| < k$.} respectively. 

Subsequently, Biely et al. proposed another link failure model that imposes an upper bound on the number of faulty links incident to each node \cite{Biely_hybrid}. As a result, it is possible to tolerate $O(n^2)$ link failures with $n$ nodes in the new model. Under this model, Schmid et al. proved lower bounds on number of nodes, and number of rounds for achieving consensus \cite{impossible_link}.
Tseng and Vaidya \cite{Tseng_netys14} considered the iterative consensus problem in arbitrary directed graphs under transient Byzantine link failure model. In particularly, they showed the tight condition on the underlying graphs for achieving iterative consensus.

For exact consensus problem, it has been shown that (i) an undirected graph of $2f+1$ node-connectivity\footnote{A graph $G=(\scriptv, \scripte)$ is said to be $k$-node connected, if $G'=(\scriptv - X, \scripte)$ is connected for all $X \subseteq \scriptv$ such that $|X| < k$.} is able to tolerate $f$ Byzantine nodes \cite{impossible_proof_lynch}; and (ii) an undirected graph of $2f+1$ edge-connectivity is able to tolerate $f$ Byzantine links \cite{Santoro_link2}.  Researchers also showed that $2f+1$ node-connectivity is both necessary and sufficient for the problem of information dissemination in the presence of either $f$ faulty nodes \cite{SS_node} or $f$ {\em fixed} faulty links \cite{SS_link}. Unlike the ``transient" failure model, the faulty links are assumed to be fixed throughout the execution of the algorithm in \cite{SS_link}. 

Charron-Bost and Schiper proposed the HO (Heard-Of) model that captures node failures and message losses at the same time \cite{HeardOf}. 
To the best of our knowledge, the HO model is the first model unifying system synchrony and node crashes together.
The HO model assumes round-based algorithms, which consists of three steps: (i) send messages, (ii) receive messages, and (iii) perform computation (specified by the algorithm). For each round $r$ and each node $i$, let $HO(i,r)$ denote the set of nodes that node $i$ has ``heard of'' at round $r$.
Then the model also specifies a set of \textit{communication predicates} over all $HO(i,r)$ to capture failures, message loss, or delayed messages.
The benefits of the HO models are: (i) it puts different types of failures in a unified framework, including static, dynamic, permanent or transient faults, and (ii)
compared with the classic fault models, the impossibility proofs and correctness proofs (for given algorithms) are in general shorter and simpler \cite{HeardOf}.
In \cite{HeardOf}, Charron-Bost and Schiper discussed communication predicates that map to classic problem specification, e.g., ``Synchronous system, reliable links, at most $f$ crash failures'' or ``Partially synchronous system, eventual reliable links, at most $f$ crash failures'', and identified the relationships among these communication predicates and solvability of consensus problems specified in the HO model with these predicates.
Subsequently, Biely et al. generalized the model to ``value faults'', which would corrupt values transmitted by nodes, and thus, can be used to capture Byzantine node and link faults \cite{HO+Byz}.

\subsection{Enriched Properties}

In addition to the termination, agreement, and validity conditions discussed in Section \ref{s:classic}, there were also researches on enriching or relaxing the correctness properties.

\paragraph{Early-Stopping Property}

In synchronous systems, an algorithm has an early-stopping property if the algorithm can terminate early if there is less than $f$ faults in an execution.
Suppose that given an execution, an actual number of faults in a system is $t$, where $t \leq f$.
It has been shown that fault-tolerant consensus cannot be achieved in $\leq t+1$ rounds using deterministic algorithms in synchronous systems \cite{AA_nancy}. That is, the lower bound of round complexity is $\min\{t+2,f+1\}$ for crash faults \cite{early_stopping_crash}, omission fault \cite{early_stopping_omission}, and Byzantine faults \cite{early_deciding_Dolev}. In \cite{early_deciding_Dolev}, Dolev and Lenzen proposed a new property, namely early-deciding, 
which requires fault-free nodes to decide early but the decided nodes may continue to  send messages  in to help other undecided nodes. They
showed that an early-deciding algorithm requires more message complexity than normal consensus algorithms \cite{early_deciding_Dolev}.
The proof consists of two parts: (i) find a ``pivotal" node that  is critical for whether  the  execution  would  result  in output $0$ or output $1$, and (ii) ensure that $\Omega(f^2)$ messages have to be exchanged in certain rounds to achieve consensus. As a result, they are able to show that for any $\min\{t+2,f+1\}$-deciding binary consensus algorithm and any $1 \leq t \leq f/2$, there is an execution such that number of faults is $t$ and fault-free nodes send at least $f^2t/44$ messages.

\paragraph{One-Step Property}
\label{s:one}

An asynchronous consensus algorithm has one-step property if in all the executions that has no contention (i.e., all fault-free nodes propose the same input value), the algorithm terminates within in one communication step. A communication step consists of three events: (i) send messages, (ii) receive messages, and (iii) perform local computation and update local state. One-step property is first proposed for crash-tolerant consensus algorithms \cite{one_step_crash} and later extended to Byzantine consensus algorithms \cite{one_step_byz,BOSCO}. These algorithms share similar structures: (i) use communication primitives to exchange values, and produce output if there is enough match, and (ii) use traditional consensus algorithms to achieve the consensus if no output is generated in the first phase. Typically, the lower bound on the number of nodes to achieve one-step property is more than the one for classic correctness properties (validity and agreement). For example, it has been shown in \cite{BOSCO} that $n > 7f$ is necessary to achieve strong one-step property, where $n$ is the number of nodes in the system.

\paragraph{Almost-Everywhere Agreement}

King and Saia studied a slightly different problem called almost-everywhere Byzantine agreement in synchronous systems with a strong adversary that corrupt nodes adaptively \cite{Multi-valued_king}. The proposed algorithm has a very small communication overhead, $\tilde{O}(n^{1/2})$ bit per node; however, it has a small error probability. Such sacrifice is necessary to overcome the lower bound proved in \cite{Lowerbound_dolev}. The core idea of the algorithm is based on iteratively performing local  elections  in  a  tournament network. To accommodate adaptive adversary, King and Saia proposed two new techniques: (i) instead of electing a node in local election, elect arrays of random number generated by some node, and (ii) use secret sharing to exchange the contents of the arrays. Such techniques may be applied to overcome adaptive adversary in other contexts. Note that their algorithm works even when input is a binary value \cite{Multi-valued_king}, whereas, multi-valued consensus algorithms achieve optimal communication complexity $O(nL)$ only when the input size $L$ is large enough.

\commentOut{++++++
	\paragraph{Other Fault Models}
	
	Consensus problems under other types of fault models have been studied.
	\cite{Sundaram_ACC,Sundaram_journal} considered consensus under malicious faults in $f$-local model
	where for each fault-free node, up to $f$ incoming nodes may become faulty.
	For broadcast problem (in which all fault-free nodes need to learn a fault-free source's input value),
	\cite{Tseng_CPA,Pelc_broadcast,Ichimura_broadcast,CPA_DISC} considered Byzantine faults in $f$-local model.
	\cite{Tseng_general,Vijay_weight_consensus} proposed generalized fault models which specified potential locations of faults 
	and can be used to capture the reliability of nodes or correlated failures.
	
	Bansal et al. \cite{Bansal_disc11} identified tight conditions for achieving exact Byzantine consensus with {\em authentication} tools in {\em undirected} graphs. Bansal et al. discovered that all-pair reliable communication is not necessary to achieve consensus when using authentication. 
	
	++++++++}
\section{Exploration of Practical Applications}
\label{s:bft}

Fault-tolerant consensus has been adopted in many practical systems.
We start with real-world systems that are designed to tolerate crash node faults, particularly, those based on two families of algorithms -- Paxos \cite{Paxos_1988} and Raft \cite{Raft}. Then, we discuss efforts on designing systems that tolerate more complex failures and BFT (Byzantine Fault-Tolerance) systems. Finally, we compare Bitcoin-related work \cite{bitcoin_Satoshi} with BFT systems and Byzantine consensus. Typically, these systems satisfy correctness (or safety) in asynchronous network; however, to ensure progress (or liveness), there must exist some time periods that enough messages are received within time. In other words, these systems satisfy safety and liveness in partially synchronous systems.

\subsection{Paxos and Raft}

Paxos \cite{Paxos_1988,Paxos_simple,fast_paxos,CMU_paxos} is the well-known family of consensus protocols tolerating crash node faults. Since Paxos was first proposed by Lamport \cite{Paxos_1988,Paxos_simple}, variants of Paxos were developed and implemented in real-world systems, such as Chubby lock service used in many Google systems \cite{Chubby,spanner}, and membership management in Windows Azure \cite{Azure}.\footnote{We would like to thank the anonymous reviewer who pointed out that Windows Azure also uses ZooKeepr to manage virtual machines \cite{Azure_Zookeeper}.} Yahoo! also developed ZaB \cite{ZaB}, a protocol achieving atomic broadcast in network equipped with FIFO channels, and used ZaB to build the widely-adopted coordination service, ZooKeeper \cite{ZooKeeper}. ZooKeeper is later used in many practical storage systems, like HBase \cite{HBase} and Salus \cite{Salus}. Recently, many novel mechanisms have been proposed to improve the performance of Paxos, including quorum lease \cite{paxos_lease}, diskless Paxos \cite{diskless_paxos}, even load balancing \cite{Paxos_egalitarian}, and time bubbling
(for handling nondeterministic network input timing) \cite{Paxos_sosp}.  
While the original Paxos \cite{Paxos_1988,Paxos_simple} is theoretically elegant, practitioners have found it hard to implement Paxos in practice \cite{Paxos_Live}. One difficulty mentioned in \cite{Paxos_Live} is that membership/configuration management is non-trivial in practice, especially, when Multi-Paxos, and disk corruptions are considered. (Multi-Paxos is a generalization of Paxos which is designed to optimize the performance when there are multiple inputs to be agree upon \cite{Paxos_Live}.)

In 2014, Ongaro and Ousterhout from Stanford proposed a new consensus algorithm -- Raft \cite{Raft}. Their main motivation was to simplify the design of consensus algorithm so that it is easier to understand and verify the design and implementation. One interesting (social) experiment by Ongaro and Ousterhout was mentioned in \cite{Raft}: ``\textit{In an informal survey of attendees at NSDI 2012, we found few people who were comfortable with Paxos, even among seasoned researchers}''. 
To simplify the (conceptual) design, Raft integrates the consensus solving part deeply with leader election protocol and membership/configuration management protocol \cite{Raft}.  
After their publication, Raft has quickly gained popularity, and been used in practical key-value store systems such as etcd \cite{etcd} and RethinkDB \cite{RethinkDB}. Please refer to their website \cite{Raft_web} for a list of papers and implementations.\footnote{Paxos has been the de facto standard of consensus algorithms for a long time \cite{Paxos_defacto}; however, we feel that it is still of interests to discuss Raft as well, as Raft has gained more and more attentions in both academia and industry \cite{Raft_web}.}

\subsection{Arbitrary State Corruption}
\label{s:asc}

Recently, researchers explored fault model beyond crash node failures. One such fault model is called Arbitrary State Corruption (ASC) \cite{yahoo_hardening,Harden,Harden2}. In the ASC fault model, the whole state of a node may transition to an arbitrary state due to incidents like bit flips or hardware error. However, the failure is not caused by a malicious adversary. Thus,
it is generally assumed that a message from a faulty node can be detected in the ASC model \cite{Harden,Harden2}. Note that the ASC model is a proper subset of Byzantine fault model, since Byzantine nodes can behave arbitrarily, including sending messages in a way that may not be detected.

Correia et al. introduced a library, PASC, which relies on different check mechanisms (e.g., CRC code) to harden crash-fault tolerant algorithms against ASC faults \cite{yahoo_hardening}. PASC does not replicate the entire node; rather, it replicates internal states of each node; thus, the overhead is moderate comparing to BFT replication (discussed in Section \ref{s:BFT}). Behrens et al. proposed a framework to harden distributed systems using arithmetic codes, which is able to detect transit and permanent hardware errors with high probability \cite{Harden2}. Subsequently, Behrens et al.\cite{Harden} observed that the technique in \cite{yahoo_hardening} does not manage memory usage efficiently, and the mechanism in \cite{Harden2} incurs large latency due to the component for encoding executions. The authors addressed the aforementioned issues, and used their technique to harden memcached \cite{memcached} with moderate overhead \cite{Harden}.

\subsection{Byzantine Fault Tolerance (BFT)}
\label{s:BFT}

Since Castro and Liskov published their seminal work PBFT (Practical Byzantine Fault-Tolerance) \cite{PBFT_1999}, significant efforts have been devoted to improving {\em Byzantine Fault-Tolerance} (BFT). There were mainly two directions of the improvements: (i) reducing the overhead like communication costs, or replication costs, and (ii) providing higher throughput or lower latency (in the form of round complexity). Generally speaking, BFT system replicates deterministic state machines over different machines (or \textit{replicas}) to tolerate Byzantine node failures. In other words, BFT systems implement the State Machine Replication systems \cite{RSM} that tolerate Byzantine faults. The main challenge is to design a system such that it behaves like a centralized server to the clients in the presence of Byzantine failures. More precisely, the system is given requests from the clients, and the goals of a BFT system are: (i) the fault-free replicas agree on the total order of the requests, and then the replicas execute the requests following the agreed order (safety); and (ii) clients learn the responses to their requests eventually (liveness). Usually, liveness is guaranteed only in the \textit{grace periods}, i.e., when messages are delivered in time. In other words, BFT systems satisfy safety and liveness in partially synchronous networks.

\paragraph{Improving Performance}

Castro and Liskov's work on Practical Byzantine Fault-Tolerance (PBFT) showed for the first time that BFT mechanism is useful in practice \cite{PBFT_1999}. PBFT requires $3f+1$ replicas, where $f$ is the upper bound on the number of Byzantine failures in the whole system. Subsequently, Quorum-based solutions Q/U \cite{QU} and HQ \cite{HQ} have been proposed, which only require one round of communication in contention-free case (when no replica fails, and the network has stable performance and no contention on the proposed input value happens) by allowing clients directly interact with the replicas to agree on an execution order. 
Such type of mechanisms reduced latency (number of rounds required) in some case, but was shown to be more expensive in other cases \cite{Zyzzyva}. Hence, Zyzzyva \cite{Zyzzyva} focused on increasing performance in failure-free case (when no replica fails) by allowing speculative operations that increase throughput significantly and adopting a novel roll-back mechanism to recover operations when failures are detected. Zyzzyva requires $3f+1$ replicas; however, a single crash failure would significantly reduce the performance by forcing Zyzzyva protocol to run inthe  slow mode -- where no speculative operation can be executed \cite{Zyzzyva}. Thus, Kotla et al. also introduced Zyzzyva5, which can be executed in fast mode even if there are crash failures, but Zyzzyva5 requires $5f+1$ replicas \cite{Zyzzyva}. Subsequently, Scrooge \cite{Scrooge} reduced the replication cost of Zyzzyva5 by requiring the participation from clients which help detect replicas' misbehaviors. 

Clement et al. observed that a single Byzantine replica or client can significantly impact the performance of HQ, PBFT, Q/U and Zyzzyva \cite{Aardvark}. Thus, they proposed a new system Aardvark, which provides good performance when Byzantine failures happen by sacrificing the failure-free case performance \cite{Aardvark}.
Later, Clement et al. also demonstrated how to combine Zyzzyva and Aardvark so that the new system, Zyzzyvark, not only tolerates faulty clients, but also enjoys fast performance in  the failure-free case by leveraging speculative operations \cite{UpRight}.

The aforementioned BFT systems are designed to optimize performance for certain circumstances, e.g., HQ for contention-free case and Zyzzyva for failure-free case.
Guerraoui et al. proposed a new type of BFT systems that can be constructed to have optimized performance under difference circumstances \cite{BFT700}. Their tunable design is useful, since it provides the flexibility of choosing different performance trade-off according to the network performance and application requirements. 
Their systems are based on three core concepts: (i) abortable requests, (ii) composition of (abortable) BFT instances, and (iii) dynamic switching among BFT instances.
The tunable parameter specifies the progress condition under which a BFT instance should not abort.
Some example conditions include contention, system synchrony or node failures. 
In \cite{BFT700}, Guerraoui et al. showed how to construct new BFT systems with different parameter; particularly, they proposed (i) \textit{AZyzzyva} which composes Zyzzyva and PBFT together to have more stable performance than Zyzzyva does and faster failure-free performance than PBFT's performance, and (ii) \textit{Aliph} which has three components: PBFT, Quorum-based protocol optimized for contention-free case, and Chain-based protocol optimized for high-contention cases without
failures and asynchrony \cite{BFT700}.

For computation-heavy workload, Yin et al. proposed to separate agreement protocol from executions of clients' requests \cite{Separating}. This separation mechanism reduces the replication cost to $2f+1$. Note that the system still requires $3f+1$ replicas to achieve agreement on the order of the clients' requests, but the executions of requests, and data storage only occur at $2f+1$ replicas. Later, Wood et al. built a system, ZZ, which reduces the replication cost to $f+1$ using virtualization technique \cite{ZZ}. The idea behind ZZ is that $f+1$ active replicas are sufficient for fault detection, and when fault is detected, their  virtualization technique allows ZZ to replace the faulty replica by waking up fresh replica and retrieving current system state with small overhead \cite{ZZ}.

\paragraph{Hardening Crash-Tolerant Systems}

Since most existing systems are designed to tolerate crash faults,
there are efforts on hardening existing crash-tolerant systems against Byzantine fault models. 
Note that systems discussed in Section \ref{s:asc} were not designed to tolerate Byzantine faults, because Arbitrary State Corruption is strictly weaker than Byzantine fault model.
Haeberlen et al. was among the first to propose using log-based detection mechanism to hardening crash-tolerant systems \cite{PeerReview}. They proposed a library called PeerReview that can be used to detect Byzantine faults, and such detection can be irrefutably linked to faulty nodes -- the identity of faulty nodes can be eventually learned by all fault-free nodes.
Unfortunately, as discovered by Ho et al. \cite{Nysiad}, PeerReview can only be used to detect a subset of Byzantine failures. Ho et al. proposed
Nysiad \cite{Nysiad}, which transforms crash-tolerant protocols to Byzantine-tolerant protocols by assigning a set of guards to verify each replica's behavior. However, Nysiad needs a logically centralized service to perform configuration change, which incurs high overhead \cite{yahoo_hardening}. 
UpRight \cite{UpRight} is an architecture which integrates BFT and crash-tolerant systems together with small overhead. UpRight has inherited ideas from three prior systems: speculative execution \cite{Zyzzyva}, robustness to clients' failure \cite{Aardvark}, and agreement/execution separation \cite{Separating}.
One novelty of UpRight is the introduction of the shim layers for clients and servers for existing existing crash-tolerant systems that can order clients' requests and verify results from servers. Clement et al. used UpRight library to make ZooKeeper \cite{ZooKeeper} tolerate Byzantine faults \cite{UpRight}.

\paragraph{Hardware-based BFT}

Different from the aforementioned software-based BFT mechanisms, researchers also proposed using trusted hardware components to reduce the replication costs or to increase performance. MinBFT \cite{MinBFT} uses trusted hardware to build an unique sequential identifier generator, which is then used to verify messages from each replica. With such scheme, MinBFT only requires $2f+1$ agreeing replicas. CheapBFT \cite{CheapBFT} relies on an FPGA-based trusted components to authenticate messages, and is able to tolerate all-but-one failures, i.e., it only requires $f+1$ replicas. Recently, Istv\'{a}n et al. proposed a novel idea of using FPGA to achieve Byzantine-tolerant consensus and atomic broadcast \cite{BFT_FPGA}. Then, they showed how to use their FPGA-based atomic broadcast to make ZooKeeper tolerate Byzantine faults with small overhead (compared to crash-tolerant one). One down side of their mechanism is that the developers need to implement an application-specific network protocol \cite{BFT_FPGA}.

\paragraph{Relaxed BFT}

Inspired by the popularity of real-world eventually consistent systems (e.g., \cite{cassandra,Dynamo}), 
researchers proposed relaxed safety and liveness properties for BFT systems. CLBFT sacrifices liveness for higher safety, i.e., tolerating more replica failures, by increasing the quorum size (in proportion of the number of replicas) \cite{NotAlwaysLive}. Zeno \cite{Zeno} chose eventual consistency to provide higher availability when network partition happens. Depot \cite{Depot} only ensures a fork-join-causal consistency (a model slightly weaker than causal consistency) to eliminate ``trust'' for safety -- a client needs to trust only  himself to ensure the safety property. Prophecy \cite{Prophecy} focuses on increasing throughput for read-heavy workloads; however, Prophecy only provides delay-once consistency (a new consistency model weaker than strong consistency \cite{Prophecy}), and relies on a trusted component to detect misbehaviors. Liu et al. proposed the concept of XFT (cross fault-tolerance), which relax the degree of fault-tolerance \cite{XFT}. Particularly, XFT is correct only when all the following conditions hold: (i) only crash faults happen in asynchronous periods; and (ii) non-crash faults (Byzantine faults) happen only in synchronous periods (grace periods).
By relaxing the guarantees, the authors build XPaxos which has comparable performance of crash-fault-tolerant systems and tolerates Byzantine faults (in grace periods) \cite{XFT}.

\paragraph{BFT Storage System}

There are also BFT systems specifically designed for storage systems. Goodson et al. proposed an erasure-coded storage system tolerating Byzantine replicas and clients using $4f+1$ replicas \cite{CMU_Erasure}. The main technique to detect faulty client writing different values to different replicas is having the next fault-free client detect the inconsistency (This scheme is possible due to the benefit of coding). Based on this system, Abd-El-Malek et al. proposed a lazy verification protocol to reduce client's workload, which shifts the work to storage replicas during idle time \cite{CMU_Lazy}. However, the scheme still requires $4f+1$ replicas, and consumes high bandwidth \cite{CMU_Erasure2}. Later, Hendricks et al. built another erasure-coded storage system \cite{CMU_Erasure2} which relies on a short checksum comprised of cryptographic hashes and homomorphic fingerprints to optimize the throughput in the contention-free case (when no replica fails, and the network has stable performance and no contention happens). The system requires $3f+1$ replicas. Recently, Cachin et al. built a BFT storage system, MDStore, which only requires $2f+1$ replicas under the assumption that the client is always fault-free when writing data \cite{IBM_Storage,Cachin2014}. MDStore system had two novelties: (i) separation of data and metadata storage, and (ii) metadata service based on lightweight cryptographic hash functions. MDStore tolerates any number of Byzantine readers and crash-faulty writers and up to $f$ Byzantine faulty replicas.

\paragraph{Cloud-of-Clouds}

The idea of building BFT storage systems over intercloud (or cloud-of-clouds) becomes popular lately, since as discussed in \cite{IBM_Intercloud}, the assumption of failure independence holds naturally due to the different cloud administrators, geographical locations and implementations from different cloud service providers. Cachin et al. proposed a layered architecture for BFT storage systems over intercloud, ICStore (abbreviating InterCloud Storage) \cite{IBM_ICStore}. 
One novelty of ICStore is to provide different dependability goals: (i) confidentiality, (ii) integrity, and (iii) reliability and consistency.
ICStore's layered architecture allows clients to choose different levels of dependability and performance by selecting different operation point for each layer \cite{IBM_ICStore}.
Independently, Bessani et al. proposed DEPSKY, a BFT storage system supporting efficient encoding and confidentiality \cite{DepSky} with $3f+1$ replicas. However, the liveness property is slightly weakened in DEPSKY, i.e., the read protocol ensures responses only when a finite number of contending writes happen. 
Lately, He et al. proposed NCCloud, which focuses on both fault tolerance and storage repair \cite{NCCloud} and is based on a new regenerating code that has low repair cost and can be used to detect a Byzantine behaviors. 

\commentOut{++++++
\paragraph{Summarization}

Here, we summarize the key contributions in BFT literatures.

\begin{center}
	\begin{tabular}{ | l | p{3cm} | l | p{2cm}| p{5cm} |}
		\hline
		{\bf System} & {\bf Num. of Replicas} & {\bf Safety} & {\bf Partition Tolerance} & {\bf Special Note} \\ \hline\hline
		
		PBFT \cite{PBFT_1999} & $3f+1$ & Strong  & No & PBFT is the first work making BFT system practical \\ \hline
		
		Zyzzyva \cite{Zyzzyva} & $3f+1$ & Strong  & No & Zyzzyva is optimized for the failure-free case by allowing speculative execution \\ \hline
		
		ZZ \cite{ZZ} & $3f+1$ agreeing replicas and $f+1$ execution replicas & Strong  & No & ZZ relies on virtualization techniques to quickly replace faulty replicas \\ \hline
		
		Aardvark \cite{Aardvark} & $3f+1$ & Strong  & No& Aardvark tolerates faulty clients but sacrifices performance in failure-free case \\ \hline
		
		CheapBFT \cite{CheapBFT} & $f+1$ active replicas for both agreement and execution & Strong  & No & CheapBFT relies on FPGA-based trusted hardware \\ \hline
		
		MDStore \cite{IBM_Storage} & $2f+1$ & Strong & No & Low replication cost in MDStore is due to (i) separation of metadata and data; and (ii) the assumption that clients always behave correctly when writing data. \\ \hline
		
		Depot \cite{Depot} & $-$ & Fork-Join-Causal  & Yes & Depot ensures safety without trusting any replica \\ \hline
		
		Zeno \cite{Zeno} & $3f+1$ & Eventual  & Yes & Zeno sacrifices safety to tolerate network partition\\ \hline

	\end{tabular}
\end{center}

+++++++}

\subsection{Bitcoin}

\textit{Bitcoin} is a digital currency system proposed by Satoshi Nakamoto \cite{bitcoin_Satoshi} and later gained popularity due to its characteristics of anonymity and decentralized design \cite{bitcoin_web}. Since Bitcoin is based on cryptography tools (Proof-of-Work mechanism), it can be viewed as a cryptocurrency.
Even though Bitcoin has large latencies (on the order of an hour), and the theoretical peak throughput is up to 7 transactions per second \cite{bitcoin_vukolic}, Bitcoin is still one of the most popular cryptocurrencies. Here, we briefly discuss the core mechanism of Bitcoin and compare it with Byzantine consensus and BFT systems.

\paragraph{Bitcoin Mechanism}

The core of Bitcoin is called {\em Blockchain}, which is a peer-to-peer ledger system, and acts as a virtually centralized ledger that keeps track of all bitcoin transactions.
A set of bitcoin transactions are recorded in blocks. 
Owners of bitcoins can generate new transactions by broadcasting signed blocks to the Bitcoin network.\footnote{Here, we follow the convention of Bitcoin literature: (i) Bitcoin network consists of all the anonymous participants in the Bitcoin system. Note that in previous sections, network means the communication network; and (ii) throughout the discussion, ``Bitcoin'' means the system, whereas, ``bitcoin'' means the virtual money.}
Then, a procedure called {\em mining} confirms the transactions and includes the transactions to the Blockchain (the centralized ledger system).
Essentially, \textit{mining} is a randomized distributed consensus component that confirms pending transactions by including them in the Blockchain. 
To include a transaction block, a miner needs to solve a ``proof-of-work'' (POW) or ``cryptographic puzzle''.
The main incentive mechanism for Bitcoin participants to maintain the Blockchain and to confirm new transactions is to reward the participants (or the miners) some bitcoins
-- the first miner that solves the puzzle receives a certain amount of bitcoins.
The main reason that the mining procedure can be related to consensus is because each miner maintains the chain of blocks (Blockchain) at local storage, and the global state is consistent at all miners eventually -- all fault-free miners will have the same Blockchain eventually \cite{bitcoin_Satoshi}. That is, anonymous Bitcoin participants need to agree on the total order of the transactions.


One important feature of the cryptocurrency system is to prevent the \textit{double-spending attacks}, i.e., spending some money twice.
In Bitcoin, the consistent global state -- the order of transactions -- can be used to prevent double-spending attacks, since the attackers have no ability to reorganize the order of blocks (i.e., modify the Blockchain, the ledger system).
In \cite{bitcoin_Satoshi}, Satoshi Nakamoto presented a simple analysis that showed with high probability, Bitcoin's participants maintain a total order of the transactions if adversary's computation power is less than $1/3$ of total computation power. As a result, no double-spending attack is possible with high probability if adversary's computation power is bounded. However, the models under consideration were not well-defined and the analysis was not rigorous in \cite{bitcoin_Satoshi}. Thus,
significant efforts have been devoted to formally proving the correctness of Bitcoin mechanism or improving the design and performance. Please refer to \cite{bitcoin_princeton} for a thorough discussion. Below, we focus on the comparison of Bitcoin and Byzantine Consensus/BFT systems.

\paragraph{Comparison with Byzantine Consensus}

There are several differences between the problem formulation of Byzantine consensus (as described in Section \ref{s:classic}) and the assumptions of Bitcoin \cite{bitcoin_garay,bitcoin_miller_anonymous,bitcoin_Satoshi}, such as in Bitcoin, (i) the number of participants is dynamic; (ii) participants are anonymous, and the participants cannot authenticate each other; (iii) as a result of (ii), participants have no way to identify the source of a received message; and (iv) the Bitcoin network is able to synchronize in the course of a round, i.e., the network communication delay is negligible compared to computation time. 

It was first suggested by Nakamoto that Bitcoin's POW-based mechanism can be used to solve Byzantine consensus \cite{bitcoin_discussion,dugcampbell}. However, the discussion was quite informal \cite{bitcoin_discussion}.
To the best of our knowledge, Miller and LaViola were the first one to formalize the suggestion and proposed a POW-based model to achieve Byzantine consensus when majority of participants are fault-free. However, the validity is only ensured with non-negligible probability (but not with over-whelming probability).
Subsequently, Garay et al. \cite{bitcoin_garay} extracted and analyzed the core mechanism of Bitcoin \cite{bitcoin_garay}, namely Bitcoin Backbone.
They first identified and formalized two properties of Bitcoin Backbone: (i) {\em common prefix property}: fault-free participants will possess a large common prefix of the blockchain, and (ii) {\em chain-quality property}: enough blocks in the blockchain are contributed by fault-free participants.
Then, they presented a simple POW-based Byzantine consensus algorithm which is a variation of Nakamoto's suggestion \cite{bitcoin_discussion}, but satisfy agreement and validity assuming that the adversary’s computation power (puzzle-solving power) is bounded by $1/3$.
Their algorithm can also be used to solve Byzantine consensus with strong validity \cite{Multi-valued_Neiger}.
Finally, they proposed a more complicated consensus protocol, which was proved to be secure assuming high network synchrony and that the adversary’s computation power is strictly less than $1/2$.
In \cite{bitcoin_garay}, Garay et al. focused on how to use Bitcoin-inspired mechanism to solve Byzantine consensus. 


\paragraph{Comparison with BFT System}

 Conceptually, BFT and Bitcoin have similar goals:

\begin{itemize}
	\item \textit{BFT}: clients' requests are executed in a total order distributively, and 
	
	\item \textit{Bitcoin}: a total order of blocks are maintained by each participant distributively.
\end{itemize}
Therefore, it is interesting to compare BFT with Bitcoin as well. Below, we address fundamental differences between the two.

\begin{itemize}
	\item {\em Formulation}: As discussed above, model assumptions for BFT are similar to the ones for Byzantine consensus, which are very different from the ones for Bitcoin. One major difference is the anonymous node identity. In BFT, the system environment is well-controlled, and replicas' IDs are maintained and managed by the system administrators. In contrast, Bitcoin is a decentralized system where all the participants are anonymous. As a result, BFT systems can use many well-studied tools from the literature, e.g., atomic broadcast, and quorum-based mechanism, whereas, Bitcoin-related systems usually rely on POW (proof-of-work) or variants of cryptographic tools.
	
	\item {\em Features}: In \cite{bitcoin_vukolic}, Marko Vukolic mentioned that the features of BFT and Bitcoin are at two opposite ends of the scalability/performance spectrum due to different application goals. Generally speaking, BFT systems offer good performance (low latency and high throughput) for small number of replicas ($\leq 20$ replicas), whereas, Bitcoin scales well ($\geq 1000$ participants), but the latency is prohibitively high and throughput is limited.
	
	\item {\em Incentive}: In BFT system, every fault-free replica/client is assumed to follow the algorithm specification. However, in Bitcoin, participants may choose not to spend their computation power on solving puzzles; thus, there is a mechanism in Bitcoin to reward the mining process \cite{bitcoin_Satoshi}.
	
	\item {\em Correctness property}: As addressed in Section \ref{s:BFT}, BFT systems satisfy safety in asynchronous network and satisfy liveness when network is synchronous enough (in grace period). As shown in \cite{bitcoin_Satoshi,bitcoin_garay}, Bitcoin requires network synchronous enough for ensuring correctness (when network delay is negligible compared to computation time).
\end{itemize}

In \cite{bitcoin_vukolic}, Marko Vukolic proposed an interesting research direction on finding the synergies between Bitcoin-related and BFT systems, since both systems have its limitations.
On one hand, the poor performance of POW-based mechanism limits the applicability of Blockchain in other domains like smart contract application \cite{bitcoin_vukolic,bitcoin_sok}. 
On the other hand, BFT systems are not widely adopted in practice due to their poorer scalability and lack of killer applications \cite{XFT,IBM_Intercloud}. SCP is a recent system that utilizes hybrid POW/BFT architecture \cite{bitcoin_SCP}. However, further exploration of the synergy between Bitcoin and BFT systems is an interesting research direction.

\section{Conclusion and Future Directions}
\label{s:future}

\subsection{Conclusion}

Fault-tolerant consensus is a rich topic. This paper is only managed to sample a subset of recent results. To augment previous surveys/textbooks on the same topic, e.g., \cite{Byz_survey,Raynal_guide,Cachin_SMR,AA_nancy,welch_book}, we survey prior works from two angles: (i) new consensus problem formulations, and (ii) practical applications. For the second part, we focus on the Paxos- and Raft-based systems, and BFT systems. We also discuss Bitcoin which has close relationship with Byzantine consensus and BFT systems.

\subsection{Future Directions}

The identified future research directions focus on one theme: {\em bridging the gap between theory and practice.}
As discussed in the first part of the paper, researchers have explored wide variety of different (theoretical) problem formulations; however, there is no consolidated or unified framework.
As a result, it is often hard to compare different algorithms and models, and it is also difficult for practitioners to decide which algorithms are most appropriate to solve their problems.
Thus, making these results more coherent and more practical (e.g., giving rule-of-thumbs for picking algorithms) would be an important and interesting task.

In the second part, we discuss the efforts of applying fault-tolerant consensus in real systems. Unfortunately, the difficulty in implementing or even understanding the consensus algorithms prevents wider applications of consensus algorithms. 
Therefore, simplifying the (conceptual) design and verifying the implementation is also a key task.
Raft \cite{Raft} is one good example of how simplified design and explanation could help gain popularity and practicability.
Another major task is to understand and analyze more thoroughly the real-world distributed systems. As suggested in \cite{bitcoin_vukolic,bitcoin_garay}, BFT systems and Bitcoin are not yet well-understood. The models presented in \cite{bitcoin_garay,bitcoin_miller_anonymous} and other works mentioned in \cite{bitcoin_vukolic} were only the first step toward this goal. Only after enough research and understanding, could we improve the state-of-art mechanisms. For example, as mentioned in \cite{bitcoin_princeton}, Bitcoin's core mechanism depends on the incentives to reward miners; however, not much work has analyzed Bitcoin from the perspective of game theory.

\section*{Acknowledgment}

We would like to thank the anonymous reviewers from SIROCCO 2016 for encouragement and suggestions. We also acknowledge Nitin H. Vaidya for early feedback and Michel Raynal for pointers to several new works.

\bibliographystyle{splncs03}
\bibliography{paperlist,newbft,BFTNitin,paperlist2,paperlist3,Bitcoin,newbft2}

\end{document}